\def\draftversion{false}
\RequirePackage{ifthen}
\ifthenelse{\equal{\draftversion}{true}}{
  \documentclass[aps,prl,10pt,galley,amsmath,amssymb, 
                 superscriptaddress]{revtex4}
}{
  \documentclass[aps,prl,10pt,twocolumn,amsmath,amssymb, 
                 superscriptaddress]{revtex4-1}
}

\ifthenelse{\equal{\draftversion}{true}}{
  \marginparwidth 2.7in
  \marginparsep 0.5in
  \newcounter{comm} 
  \def\commnext{\stepcounter{comm}}
  \def\commtext{{\bf\color{blue}[\arabic{comm}]}}
  \def\commmar{{\bf\color{blue}[\arabic{comm}]}}
  \def\dvm#1{\commnext\marginpar{\small DV\commmar: #1}\commtext}
  \def\khm#1{\commnext\marginpar{\small KH\commmar: #1}\commtext}
  \def\hsm#1{\commnext\marginpar{\small HS\commmar: #1}\commtext}
  \def\mlab#1{\marginpar{\small\bf #1}}
  
}{
  \def\dvm#1{}
  \def\khm#1{}
  \def\hsm#1{}
  \def\mlab#1{}
  
}

\usepackage{soul}  

\usepackage[]{cleveref}
\usepackage[]{graphicx}
\usepackage[]{verbatim}
\usepackage[dvipsnames]{xcolor}
\usepackage{tabularx}

\usepackage{amsfonts}
\usepackage{amsmath}
\usepackage{amssymb}
\usepackage{bm}
\usepackage{graphicx}
\usepackage[breaklinks,colorlinks=true,citecolor=blue]{hyperref}
\usepackage{mathrsfs}
\usepackage[lofdepth,lotdepth,caption=false]{subfig}
\usepackage{varwidth}
\usepackage{wrapfig}
\usepackage{times}
\usepackage{longtable}
\usepackage{multirow}
\usepackage{tikz}

\makeatletter

\AtBeginDocument{\@ifpackageloaded{natbib}{\ifNAT@numbers\if@filesw\immediate\write\@auxout{\string\global\string\NAT@numberstrue}\fi\fi}{}}
\makeatother
\begin{document}

\title{
 Mott metal-insulator transitions in pressurized layered
trichalcogenides
}

\author{Heung-Sik Kim}
\affiliation{Department of Physics and Astronomy, Rutgers University,
Piscataway, New Jersey 08854-8019, USA}
\affiliation{Department of Physics, Kangwon National University,
Chuncheon 24341, Korea}

\author{Kristjan Haule}
\affiliation{Department of Physics and Astronomy, Rutgers University,
Piscataway, New Jersey 08854-8019, USA}

\author{David Vanderbilt}
\affiliation{Department of Physics and Astronomy, Rutgers University,
Piscataway, New Jersey 08854-8019, USA}

\begin{abstract}
Transition metal phosphorous trichalcogenides, $M{\rm P}X_3$ ($M$ and
$X$ being transition metal and chalcogen elements respectively), have
been the focus of substantial interest recently
because they are unusual candidates undergoing Mott transition 
in the two-dimensional limit. Here we investigate material
properties of the compounds with $M$ = Mn and Ni employing {\it
ab-initio} density functional and dynamical mean-field calculations,
especially their electronic behavior under external pressure in the
paramagnetic phase. Mott metal-insulator transitions (MIT) are found to
be a common feature for both compounds, but their lattice
structures show drastically different behaviors depending on the
relevant orbital degrees of freedom, {\it i.e.} $t_{\rm 2g}$ or
$e_{g}$. 
Under pressure MnPS$_3$ undergoes an isosymmetric structural transition in 
monoclinic space group by forming Mn-Mn dimers due to the strong direct overlap 
between the neighboring $t_{\rm 2g}$ orbitals, accompanied by a significant volume 
collapse and a spin-state transition. 
In contrast, NiPS$_3$ and NiPSe$_3$,
with their active $e_g$ orbital degrees of freedom, do not show
a structural change at the MIT pressure or deep in the metallic
phase within the monoclinic symmetry. 
Hence NiPS$_3$ and NiPSe$_3$ become rare examples of materials hosting
electronic bandwidth-controlled Mott MITs, thus showing promise for ultrafast
resistivity switching behavior.
\end{abstract}

\maketitle

Since the first identification of the Mott metal-insulator transition (MIT) by Mott and Peierls in 1937~\cite{Mott1937} and the suggestion of the canonical Hubbard model in 1963~\cite{Hubbard1963}, many systems showing the Mott MIT have been found. They can be broadly classified into two categories, {\it a}) the filling-controlled MITs, such as in the doped cuprates~\cite{Bednorz1986}, or {\it b}) the bandwidth-controlled MITs, such as in the rare-earth nickelates $R$NiO$_3$ ($R$ being a rare-earth element)~\cite{RNOPD,RNOreview} or vanadium oxides V$_2$O$_3$~\cite{Morin1959,McWhan1971,Carter1993} and VO$_2$~\cite{Morin1959,Qazilbash2007}. 
Considering applications to electronic resistive switching devices, the filling-controlled MIT of type ({\it a}) is not favorable due to the inevitable strong inhomogeneity at the atomic scale introduced by the chemical doping. The bandwidth-controlled MITs of type (b), on the other hand, are typically coupled strongly to the structural degrees of freedom, as for examples in the bond disproportionation between the short and long Ni-O bonds in $R$NiO$_3$~\cite{Lucian2017,Johnston2014PRL,Park2012PRL,Mizokawa2000PRB} and the dimerization of vanadium atoms in VO$_2$~\cite{Brito2016,Biermann2004}. Such involvement of slow lattice dynamics in the MIT is also not favorable for fast switching. Hence, systems with {\it electronic} bandwidth-controlled MITs ({\it i.e.} weak or no lattice distortions involved) are desirable for fast resistive switching~\cite{Ramanathan2010,Zheng2011ARMR}. 

Surprisingly, there are very few solids that are known to undergo purely electronic and bandwidth-controlled MITs, as was originally envisioned by Hubbard. From the theoretical side, there is growing evidence that starting from the metallic side, the MIT would not have occurred in any of above three systems ($R$NiO$_3$, V$_2$O$_3$, and VO$_2$) in the absence of a simultaneous structural distortion. Even less common are such transitions in two-dimensional materials, which might be useful for ultrathin electronic and spintronic applications, and to our knowledge there is no known example of an electronically driven MIT among the van der Waals (vdW) materials except the recently discovered Mott phase and superconductivity in twisted bilayer graphene~\cite{TWG_Nature2018_1,TWG_Nature2018_2}.

\begin{figure*}
  \centering
  \includegraphics[width=1.00\textwidth]{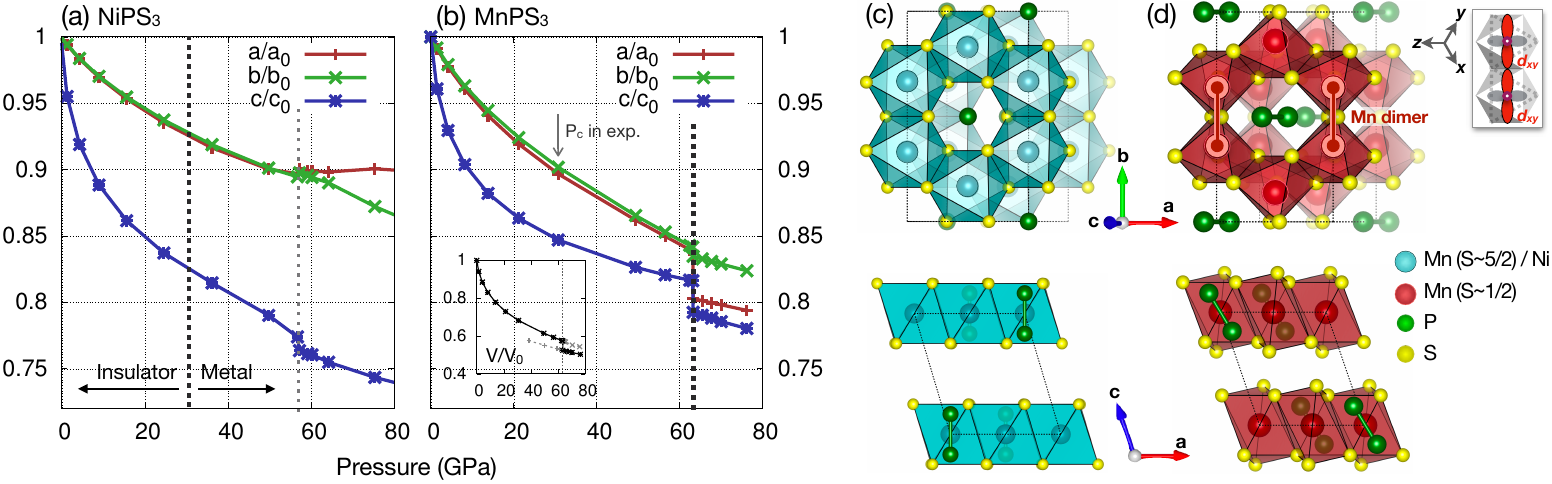}
  \caption{
  (a,b) Evolution of DFT+$U$-optimized lattice parameters 
($a$, $b$, and $c$ as depicted in (c,d)) as a function of pressure,
where (a) and (b) panels show results from NiPS$_3$ and MnPS$_3$ respectively.
Inset in (b) shows a volume versus pressure plot
for MnPS$_3$,
where black and gray curves represent ground-state and metastable
structures respectively.
Thick vertical dashed lines in both plots indicate the
values of critical pressure where the MIT happens. The thin dotted line
in (a) shows the pressure where the $a/a_0$ and $b/b_0$ begin to branch
in NiPS$_3$ ({\it i.e.} $b \neq \sqrt{3}a$) due to the enhanced
monoclinicity by pressure. Note that, the critical pressure for
MnPS$_3$ reported in Ref. \onlinecite{WangJACS2016} is around 30 GPa,
as depicted in the figure. (c) Crystal structures for NiPS$_3$ and
MnPS$_3$ at the ambient pressure. (d) MnPS$_3$ structure when P $>$ 64
GPa, where the Mn-dimer is formed parallel to $b$.
  }
  \label{fig:struct}
\end{figure*}

Here we propose new candidates for the electronic bandwidth-controlled
MIT without significant structural distortion among the emerging class of
two-dimensional vdW materials. Our target is a series of transition
metal phosphorous trichalcogenides $M{\rm P}X_3$ ($M$ = Mn, Ni, $X$ =
S, Se)~\cite{WangJACS2016,Soyeun2018,JGPark2016a}. To incorporate the electronic and structural degrees of freedom
on an equal footing, we employ the state-of-the-art embedded dynamical
mean-field theory combined with density functional theory (eDMFT),
which implements forces on atoms and allows relaxation of internal
atomic coordinates~\cite{Haule2017}. For the optimization of the size
and shape of unit cells we use density functional theory (DFT)
augmented by the on-site Coulomb repulsion $U$ (DFT+$U$), after which
optimizations of internal atomic coordinates are performed both in
eDMFT and DFT+$U$ yielding consistent results\footnote{For DFT+DMFT
calculations we use the Rutgers DFT+Embedded DMFT
code~\cite{Haule2010}, while for DFT and DFT+U calculation we used the
Vienna {\protect\it Ab-initio} Simulation Package~\cite {VASP1,VASP2}.
Dependency to different choices of exchange-correlation functionals, including
van der Waals functionals, was also checked.
For computational details refer to the Supplementary Material (SM), where we
discussed the the difference between
the $U$-values employed in our eDMFT calculations and other studies\cite{HaulePRB2014,Mandal2019}
}.
We mainly focus on paramagnetic phases of MnPS$_3$ and NiP\{S,Se\}$_3$ above their
N\'{e}el temperatures ($T_{\rm N}$ = 78 and 154 K for MnPS$_3$ and
NiPS$_3$ respectively~\cite{BREC1986,Wildes2006,Wildes2015}), with
disordered local Mn$^{2+}$ ($d^5$) $S = 5/2$ and Ni$^{2+}$ ($d^8$) $S =
1$ moments, although the behavior of their MITs in
the magnetic phases is discussed in the Supplementary Material (SM).
We will show that the recently discovered MIT in
MnPS$_3$ falls under the family of transitions
coupled to structural changes, in which the dimerization plays a
crucial role, therefore bearing a resemblance to the MIT in VO$_2$~\cite{WangJACS2016}. On
the other hand, theoretical simulations in NiPS$_3$ and NiPSe$_3$
suggest that the MIT in these two vdW compounds occurs at even lower
pressure, and does not involve a simultaneous structural transition.
Therefore they become rare examples of electronic
bandwidth-controlled transitions with a potential for very fast
resistive switching.

{\color{NavyBlue}\it Crystal structures versus pressure.}
Fig.~\ref{fig:struct}(a) and (b) show DFT+$U$ results on the
pressure-induced change of the three lattice parameters ($a/a_0$,
$b/b_0$, and $c/c_0$, where $\{a,b,c\}_0$ denote their zero-pressure
values) for NiPS$_3$ and MnPS$_3$, respectively.
Note that here we focus on the monoclinic $C2/m$ structure as shown 
in Fig. \ref{fig:struct}(c). 
Because both of the compounds are vdW-type layered systems, the
inter-plane lattice parameter $c$ shows a steeper decrease compared to
the in-plane $a$ and $b$, and the three-fold symmetry within each layer
forces $b \simeq \sqrt{3}a$ in the low-$P$ regime. The resulting volume
decrease under pressure is substantial: 40\% of volume reduction
at $\sim\!50\,$GPa compared to the ambient pressure volume, as shown in the
inset of Fig.~\ref{fig:struct}(b).

Both compounds show MIT and structural phase transitions under
pressure, but the nature of their transition is drastically
different. As shown in Fig.~\ref{fig:struct}(a), the MIT and the
structural transition in NiPS$_3$ occur at very different pressures,
around 31 and 57 GPa respectively, while they coincide in MnPS$_3$.
Remarkably, theoretical simulations suggest that the MIT in NiPS$_3$
accompanies no significant structural distortion (discontinuous
structural changes, for example), and is thus a rare example of
an electronically driven bandwidth-controlled MIT. On the other hand,
in MnPS$_3$ the isosymmetric structural transition
({\it i.e.}, structural transition within the same space group symmetry)
with a volume collapse at $63\,$GPa is crucial for the occurrence of
the MIT, hence the transition is better classified as the structurally
assisted MIT (see Fig.~\ref{fig:struct}(b)). We note that the
theoretical critical pressure of $63~\,$GPa is somewhat overestimated
compared to the experimentally reported value of
$\sim\!30\,$GPa~\cite{WangJACS2016}. However, we show in the SM that within
eDMFT, spinodal lines extend down to a much lower pressure of $~40\,$GPa
with a much reduced energy barrier between the metallic and insulating
solutions compared to the DFT+$U$ results. Inclusion of the phonon free
energy and the lattice zero-point energy, which is neglected here,
could then move the position of the transition significantly (see the SM for
further details).

In addition to the volume collapse, mostly from the discontinuous
change of $a$, DFT+$U$ simulations of MnPS$_3$ show a Mn-Mn
dimerization along the $b$-direction with the tilting of the P$_2$
dimer as shown in Fig.~\ref{fig:struct}(d). The Mn-Mn bond lengths
between the dimer and non-dimer bonds are 2.42 and 3.10~\AA~at 63 GPa,
respectively, which is a rather large difference. This Mn dimer
formation is attributed to the direct $d$-$d$ overlap between the Mn
$t_{\rm 2g}$ orbitals, pointing directly towards the nearest-neighbor
Mn as shown in the inset of Fig.~\ref{fig:struct}(d). Note that the
previous experimental study suggested the formation of Mn zigzag chains
in the high-P phase~\cite{WangJACS2016}, in contrast to our
DFT+$U$ and eDMFT results.

NiPS$_3$, on the other hand, shows no such intermetallic dimerization
or chain formation at the MIT or beyond the structural transition
pressure, because the partially-filled Ni $e_{\rm g}$ orbitals point
towards the S atoms. This makes NiPS$_3$ more sensitive to the $p$-$d$
hybridization, yielding a smaller MIT pressure in NiPS$_3$ compared to
MnPS$_3$. Such a stark contrast between NiPS$_3$ and MnPS$_3$,
originating from the difference in their orbital physics, affects the
nature of the structural behavior of their MIT as shown below.
Note that the structural transition in NiPS$_3$ at 57
GPa is not orbital in nature and comes purely from the reduced
interlayer distance and the large overlap between the layers.

We note that the two compounds show markedly different pressure
dependence of the lattice parameters even before the structural
transition. By comparing Fig.~\ref{fig:struct}(a) and (b) it is
evident that the compression of $c$ under pressure is stronger in
NiPS$_3$ than in MnPS$_3$; while $a/a_0 - c/c_0$ in MnPS$_3$ at 60 GPa
is about 0.03 (See Fig.~\ref{fig:struct}(b)), in NiPS$_3$ it is about
0.15 (Fig.~\ref{fig:struct}(a)) despite the similar volume change. In
other words, it is much easier to compress NiPS$_3$ along the
layer-normal direction compared to MnPS$_3$. Because the kinetic
energy scale set by the hopping integrals between the $t_{\rm 2g}$ (for MnPS$_3$) 
and $e_{\rm g}$ (for NiPS$_3$) orbitals show different
anisotropy, the $t_{\rm 2g}$ and $e_{\rm g}$ yield strong in-plane
$d$-$d$ and inter-plane $d$-$p$-$p$-$d$ overlaps respectively (See SM).
As a result, while the $t_{\rm 2g}$ orbitals favor in-plane compression
for the larger in-plane kinetic energy gain, the $e_{\rm g}$ orbitals
prefer to reduce the inter-plane distance, yielding
the tendency shown in Fig.~\ref{fig:struct}(a) and (b).

\begin{figure}
  \centering
  \includegraphics[width=0.80\columnwidth]{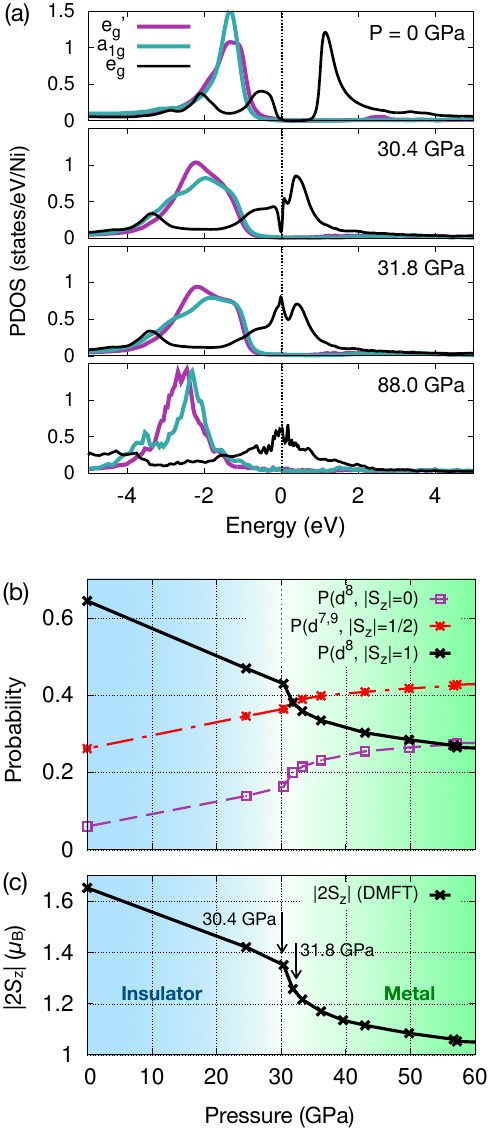}
  \caption{
  (a) Projected density of states (PDOS) in the paramagnetic phase of NiPS$_3$ at $T$ = 232K, 
  calculated by eDMFT with the increasing pressure from the ambient
condition (top panel) to 88 GPa (lowest). (b) Monte Carlo probabilities
for the $d^8$ $\vert S_{\rm z} \vert$ = 0 (purple dashed line),
$d^{7,9}$ $\vert S_{\rm z} \vert$ = 1/2 (red dash-dotted), and $d^{8}$
$\vert S_{\rm z} \vert$ = 1(black solid) as a function of pressure. (c)
Pressure dependence of the size of PM Ni spin moment $\vert 2S_{\rm z}
\vert$ from PM eDMFT results at $T$ = 232K. Note a cusp at P = 30.4 GPa
where the MIT happens.
  }
  \label{fig:NPS}
\end{figure}

{\color{NavyBlue}\it Electronic MIT in NiPS$_3$.}
Below we take a closer look into the nature of the MIT in NiPS$_3$.
Note that all the spectra presented hereafter are eDMFT results, where
the DFT+$U$-optimized cell parameters and estimated pressure values are
employed. Fig.~\ref{fig:NPS}(a) shows projected densities of states (PDOS) of
NiPS$_3$ with varying pressure from 0 to 88 GPa. 
It is clear that the $t_{\rm 2g}$ states ($a_{\rm g}$ and $e'_{\rm g}$) are mostly
occupied, while the $e_{\rm g}$ states are partially filled and show
a narrow dip at the Fermi level at 30.4 GPa. The self-energies of the $e_{\rm g}$
orbitals show poles at the Fermi level (see SM), confirming the 
presence of the paramagnetic Mott phase. 
Previously, it was suggested that NiPS$_3$ is a negative
charge-transfer (NCT) insulator with a $d^9\underline{L}^1$
configuration ($\underline{L}$ denoting a S $p$-ligand
hole)~\cite{Soyeun2018}.
However, our eDMFT results show that when the Ni occupation is close to
$n_d\approx 9$, where the $d^9\underline{L}^1$ configuration is
dominant, the Mott insulating state cannot be stabilized, i.e., the
material is metallic. The experimentally observed Mott insulating
behavior can only be achieved with the Ni occupancy of $n_d\approx 8$,
where the high-spin $S$ = 1 configuration is dominant, i.e.,
corresponding to approximately half-filled $e_{\rm g}$ states (see
Fig.~\ref{fig:NPS}(b) for the probability distribution in the
insulating and metallic states). This observation is corroborated by
X-ray absorption spectroscopy, indicating that NiPS$_3$ is close to the
NCT regime, but is still dominated by the $d^8$ $S$ = 1 configuration,
consistent with our eDMFT results~\cite{Banabir2018}.

Fig.~\ref{fig:NPS}(b) shows the valence histogram for the few most
important Ni $d$ configurations versus pressure at $T$ = 232K. The Mott
insulating state is stable as long as the high-spin state ($\vert
S_{\rm z} \vert$ = 1) of the Ni-$d^8$ configuration is dominant.
Note that we report $S_z$ values rather than $S$ values, because of our
choice of an Ising-type approximation of the Coulomb interaction in the
eDMFT impurity solver~\footnote{This approximation leads to some mixing
between $S$ = 0 and 1 states, but is not expected to change qualitative
aspects of the results}.
 Around 31$\,$GPa the $\vert S_{\rm z} \vert$ = 1/2 states (of $d^9$
and $d^7$ configurations) become equally probable, at which point the
Mott state collapses and a narrow metallic quasiparticle peak appears
(see the third panel in Fig.~\ref{fig:NPS}(a)). Despite the enhanced
charge fluctuation, the change of Ni $d$-orbital occupation ($n_d$)
across the transition is negligible: $n_d$ = 8.15 and 8.19 at P = 0 and
88 GPa, respectively.
The increase of charge fluctuations with increasing pressure has an
additional effect of unlocking the $\vert S_{\rm z} \vert$ = 0
sector of the $d^8$ configuration, which is favored in the itinerant low spin regime at
large pressure. Note that the increase of the probability for $\vert
S_{\rm z} \vert$ = 0 at the expense of the $\vert S_{\rm z} \vert$ = 1
state has a large effect on the size of the fluctuating moment $\vert
2S_{\rm z}\vert$, which is plotted in Fig.~\ref{fig:NPS}(c). Its
zero-pressure value is around $1.6\,\mu_B$, which is quite reduced from the
maximum atomic value of $2\,\mu_B$, and once it is reduced below
$1.4\,\mu_B$ it drops very suddenly and takes values of $\le 1.3\,\mu_B$ in
the metallic state. We note that the change between 30.4 and
31.8$\,$GPa is abrupt, which is likely associated with a first-order
transition, for which a coexistence of both solutions is expected.
However the hysteresis was not observed, probably because
it is too narrow at the temperature studied ($T=232\,$K).
We mention that the MIT was carefully checked by employing
beyond-Ising Coulomb interaction terms (spin-flip and pair-hopping types) at a 
lower temperature of $T=116\,$K, but neither a hysteretic behavior nor
a discontinuity in the energy-volume curve were found, signifying 
very weak coupling between the lattice and charge degrees of 
freedom in NiPS$_3$ (see Sec. III. A in the SM and Fig. S3 therein for more details).


\begin{figure}
  \centering
  \includegraphics[width=0.96\columnwidth]{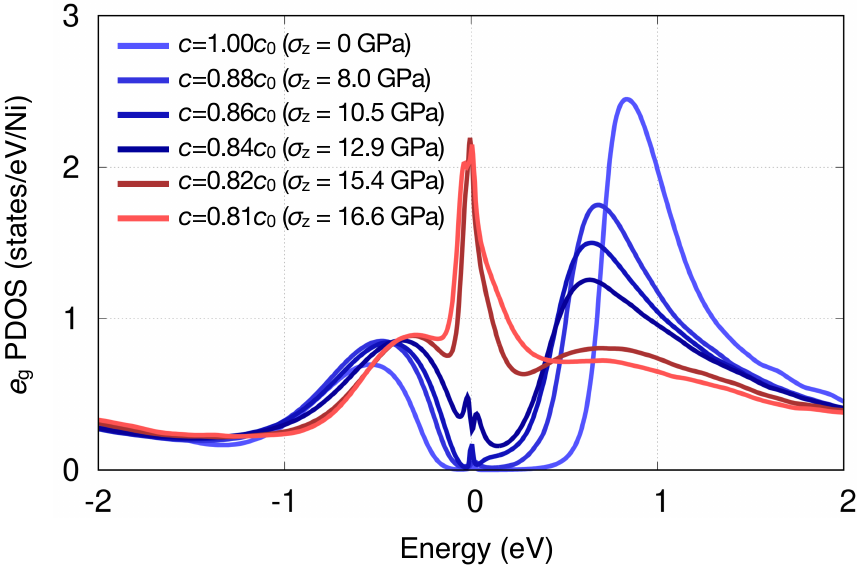}
  \caption{
NiPSe$_3$ $e_{\rm g}$-PDOS in the presence of an uniaxial stress
$\sigma_z$ along the layer-normal $\hat{z}$-direction from PM eDMFT
results at $T$ = 58 K (with varying $\sigma_z$ from 0 to 16.6 GPa). Blue and red
curves are PDOS for insulating and metallic phases, respectively, when
the MIT happens between $\sigma_z$ = 15.4 and 16.6 GPa.
  }
  \label{fig:NPSe}
\end{figure}

{\color{NavyBlue}\it MIT driven by uniaxial pressure in NiPSe$_3$.}
While the critical pressure for the MIT in NiPS$_3$ can be reached in
modern high-pressure experimental setups, a substitution of S by the more
polarizable Se is expected to further reduce the critical pressure.
Therefore the recently synthesized NiPSe$_3$~\cite{Yan2017PRM} 
can be a better candidate for realizing the pressure-driven MIT 
compared to NiPS$_3$. Moreover, the
collapse of the interlayer distance is expected to be sufficient to
induce the MIT, which can even be achieved by the tip of an
atomic-force microscope~\cite{Lu59}. As a zeroth-order approximation,
we simulate such layer-normal strain by varying the interlayer distance
with fixed in-plane lattice parameters, and allowing the internal
coordinates to relax within eDMFT. In Fig.~\ref{fig:NPSe} we show the
$e_{\rm g}$-PDOS of NiPSe$_3$ where the MIT happens at the modest
stress of $13 \leq \sigma_z \leq 15$ GPa at $T$ = 58 K,
%
%
suggesting NiPSe$_3$ as another promising
electronic bandwidth-controlled Mott transition system
among these layered vdW materials.

\begin{figure}
  \centering
  \includegraphics[width=0.8\columnwidth]{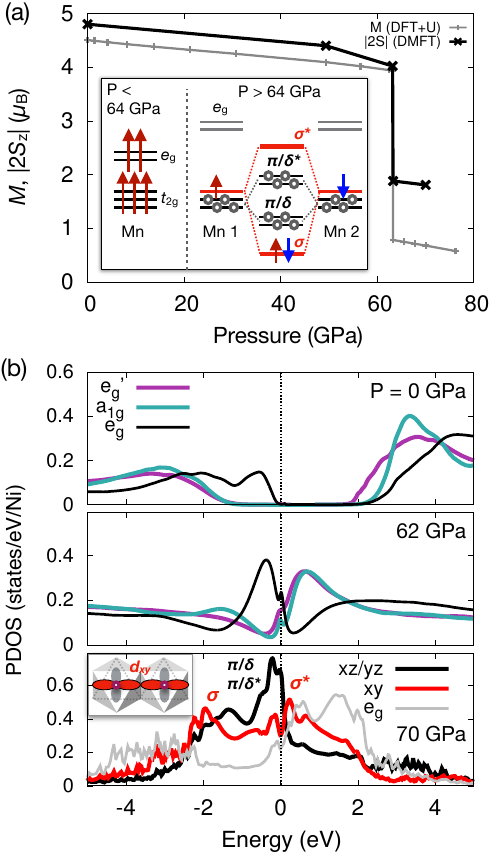}
  \caption{
    (a) Pressure dependence of the size of Mn spin moment M from
DFT+$U$ (gray line) and $\vert 2S_{\rm z} \vert$ from paramagnetic
eDMFT at $T$ = 580K (black), where both results show the spin-state
transition at 63 GPa. The schematic spin-orbital configurations below
and above the transition are shown in the inset. (b) PDOS from eDMFT,
at P = 0 GPa (upper panel), 62 GPa (center), and 70 GPa (lower). Note
that the choice of $d$-orbital onto which the DOS are projected is
different below and above the transition ($e'_{\rm g}$ and $a_{\rm 1g}$
below, and $d_{xz,yz,xy}$ above 63 GPa).
  }
  \label{fig:MPS}
\end{figure}

{\color{NavyBlue}\it Volume collapse and MIT in MnPS$_3$.} We now
address the volume-collapse transition in MnPS$_3$. We first
checked that once the optimized lattice parameters from DFT+$U$ are
employed, both DFT+$U$ and eDMFT optimizations for the
internal coordinates yield practically the same result.
Fig.~\ref{fig:MPS}(a) shows the evolution of the fluctuating Mn moment
$\vert 2S_{\rm z} \vert$ within the PM eDMFT at $T$ = 580 K. We also
show the ordered magnetic moment $M$ (per Mn) within DFT+$U$ in the
N\'{e}el-type antiferromagnetic ordered
state~\cite{BREC1986,Okuda1986,WangJACS2016} (grey line in the plot).
Perhaps not surprisingly, the two methods show very similar behavior:
an insulating state of almost maximum spin $S$ = 5/2 configuration in
the low-to-intermediate-P regime, and a metallic state with strongly
reduced Mn moments above P = 64 GPa. The inset of Fig.~\ref{fig:MPS}(a)
schematically depicts the spin-orbital configuration below and above
the transition. In the presence of strong external pressure, the
orbitally inert high-spin $S$ = 5/2 configuration becomes energetically
unstable, and the low-spin $S$ = 1/2 with the partially filled $t_{\rm
2g}$ orbital is stabilized with an octahedral volume
collapse~\cite{Kunes2008}. As a result, the $t_{\rm 2g}$ open shell in this
edge-sharing geometry leads to a strong $\sigma$-like direct $d$-$d$
overlap between the nearest-neighboring (NN) Mn sites, resulting in a
strong tendency toward Mn dimerization~\cite{Streltsov2016}. 
At ambient pressure, because of the weak inter-layer coupling, 
the three in-plane NN bonds are essentially equivalent to each
other. In the high-pressure regime, however, the monoclinicity 
originating from the layer stacking is no longer negligible. 
Therefore, the NN bond parallel to the $b$ direction, which becomes 
nonequivalent to the other bonds, dimerizes as shown in
Fig.~\ref{fig:struct}(d)
(see Sec. III. B in the SM for more details on inclusion of the beyond-Ising Coulomb terms).

{\color{NavyBlue}\it Summary.} We report a theoretical study of the Mott MIT induced by external pressure in strongly correlated layered vdW materials. 
We comment that the Mott phases in other metal trisulfides, such as FePS$_3$~\cite{JGPark2016b} or CoPS$_3$~\cite{WIldes2017JPCM}, are also of great interest because of their partially filled $t_{\rm 2g}$ shells even under ambient conditions. Overall, this family of vdW-layered transition metal trichalcogenides can be an excellent platform for the study of strong electron correlations and their cooperation with spin and lattice degrees of freedom.

\begin{acknowledgments}
{\color{NavyBlue}\it Acknowledgments:} We thank Matthew J. Coak, 
Michael O. Yokosuk, Nathan C. Harms, Kevin A. Smith, Sabine N. Neal, 
Janice L. Musfeldt and Sang-Wook Cheong for helpful discussions.
The work was supported by NSF DMREF grant DMR-1629059.
HSK thanks the support of the 2019 research grant for 
new faculty members from Kangwon National University, and also
the support of supercomputing resources including technical assistances 
from the National Supercomputing Center of Korea 
(Grant No. KSC-2019-CRE-0036)
\end{acknowledgments}

\bibliography{MPS3}{}

\clearpage

\renewcommand{\thefigure}{S\arabic{figure}}

\section{Computational details}
\subsection{Density functional theory calculations}

For unit cell optimizations and relaxations of initial internal coordinates, the Vienna {\it ab-initio} Simulation Package ({\sc vasp}), which employs the projector-augmented wave (PAW) basis set~\cite{VASP1,VASP2}, was used for density functional theory (DFT) calculations in this work. 340 eV of plane-wave energy cutoff and 8$\times$6$\times$8 Monkhorst-Pack $k$-grid sampling were employed. For the treatment of electron correlations within DFT, a revised Perdew-Burke-Ernzerhof exchange-correlation functional for crystalline solid (PBEsol) was employed\cite{PBEsol}, in addition augmented by on-site Coulomb interactions for transition metal $d$-orbitals within a simplified rotationally-invariant form of DFT+$U_{\rm eff}$ formalism\cite{Dudarev1998PRB}. $10^{-4}$ eV/\AA~of force criterion was employed for structural optimizations. For test purpose, Ceperley-Alder local density approximation\cite{LDA-CA} and the original PBE\cite{GGA-PBE} functionals were also used. 

Structural relaxations for all compounds were performed in the presence of the DFT+$U_{\rm eff}$ (4 eV) on-site Coulomb interaction and a N\'{e}el-type antiferromagnetic order\cite{BREC1986}, which gives reasonable agreements of lattice parameters and gap sizes with experimentally observed values\cite{WangJACS2016,Soyeun2018}. It should be mentioned that, without incorporating magnetism and $U_{\rm eff}$ to open the gap, the volume is severely underestimated for both compounds, especially $\sim$ 20\% smaller in MnPS$_3$. This observation signifies the role of electron correlations in structural properties of these compounds. 

\subsection{Dynamical mean-field theory calculations}
A fully charge-self-consistent dynamical mean-field method\cite{Haule2010}, implemented in Rutgers DFT + Embedded DMFT (eDMFT) Functional code (\href{http://hauleweb.rutgers.edu/tutorials/}{http://hauleweb.rutgers.edu/tutorials/}) which is combined with {\sc wien2k} code\cite{wien2k}, is employed for computations of electronic properties and optimizations of internal coordinates\cite{Pascut2016}. In DFT level the Perdew-Wang local density approximation (LDA) is employed, which was argued to yield the best agreement of lattice properties combined with DMFT\cite{Haule2015FE}. 500 $k$-points were used to sample the first Brillouin zone with $RK_{\rm max}$ = 7.0. A force criterion of 10$^{-4}$ Ry/Bohr was adopted for optimizations of internal coordinates. A continuous-time quantum Monte Carlo method in the hybridization-expansion limit (CT-HYB) was used to solve the auxiliary quantum impurity problem\cite{HauleQMC}, where the full 5 $d$-orbitals of Ni and Mn were chosen as our correlated subspaces in a single-site DMFT approximation. For the CT-HYB calculations, up to 10$^{10}$ Monte Carlo steps were employed for each Monte Carlo run. In most runs temperature was set to be 232K, but in some calculations with high pressure it was increased up to 580K because of the increased hybridization between the impurity and bath. -10 to +10 eV of hybridization window (with respect to the Fermi level) was chosen, and $U$ = 10 eV and $J_{\rm H}$ = 1 eV of on-site Coulomb interaction parameters were used for both Mn and Ni $d$-orbitals. A simplified Ising-type (density-density terms only) Coulomb interaction was employed in this work, and it was tested that the use of full Coulomb interaction yields only quantitatively different results in terms of pressure-induced evolution of electronic structures; see Sec. III for more details. A nominal double counting scheme was used, with the $d$-orbital occupations for double counting corrections for Ni and Mn were chosen to be 8 and 5, respectively.  
%
%

We comment that the choice of optimal values of the Coulomb interaction $U$ is method-dependent.
Other than the eDMFT approach chosen in this study, there are two widely employed first-principles methods using $U$ to incorporate electron correlations; (a) DFT+$U$, and (b) DFT+DMFT with Wannierized correlated orbitals. Both methods use smaller values of $U$ ($\simeq$ 4 eV) for the correlated $d$ orbitals in transition-metal compounds compared to eDMFT (10 eV)\cite{HaulePRB2014,Mandal2019}. 
First, unlike in DFT+$U$, in DFT+DMFT formalisms all (local) dynamic screening processes are included via exactly solving the many-body impurity problem. Being such screening processes explicitly treated within DFT+DMFT means that, the input Coulomb interaction $U$ should be closer to the bare one (only screened by the core and semi-core states). In DFT+$U$, on the contrary, one should use the screened $U$ (whose value smaller than the DMFT $U$) to compensate the missing screening processes therein. Hence the $U$ values employed in our eDMFT results are larger than the values used in DFT+$U_{\rm eff}$ calculations. 

Secondly, for DFT+DMFT with Wannierized correlated orbitals, it is well known that the Wannier functions contain a substantial amount of $p$ character from anions (oxygen or chalcogen ions) if a narrow Wannierization energy window that contains only the correlated subspace is chosen. Consequently the critical $U$ becomes much smaller because of the mixing of $p$ character, to be indeed of the order of the bandwidth $W$. Note that this approach is equivalent to solving Hubbard-type models, where only correlated orbital degrees of freedom are considered. 

Contrary to the aforementioned approaches, in the eDMFT formalism we are solving a generalized Anderson-lattice type Hamiltonian (actually the $p$-$d$ type Hamiltonian), where the effective $U_{\rm eff}$ that could be compared with the Hubbard-$U$ in the Hubbard-type model is actually the $p$-$d$ splitting. The advantage of using such $p$-$d$ type Hamiltonian in DFT+DMFT is evident; the $U$ values in such models are much more system-independent for many transition-metal compounds, as demonstrated recently\cite{HaulePRB2014,Mandal2019}. Therein we established that a reasonable $U$ for a wide range of transition-metal oxides within the eDMFT ({\it i.e.} the $p$-$d$ model with very localized $d$ orbitals described above) is around 10 eV and $J$ = 1 eV, and is much more universal than the $U$ values in downfolded Hubbard-like models.

\section{Exchange-correlation and van der Waals functional dependence within DFT+$U_{\rm eff}$}

\subsection{Dependence on exchange-correlation functionals and $U_{\rm eff}$-value in DFT+$U_{\rm eff}$ results}

\begin{figure}
\includegraphics[width=0.40\textwidth]{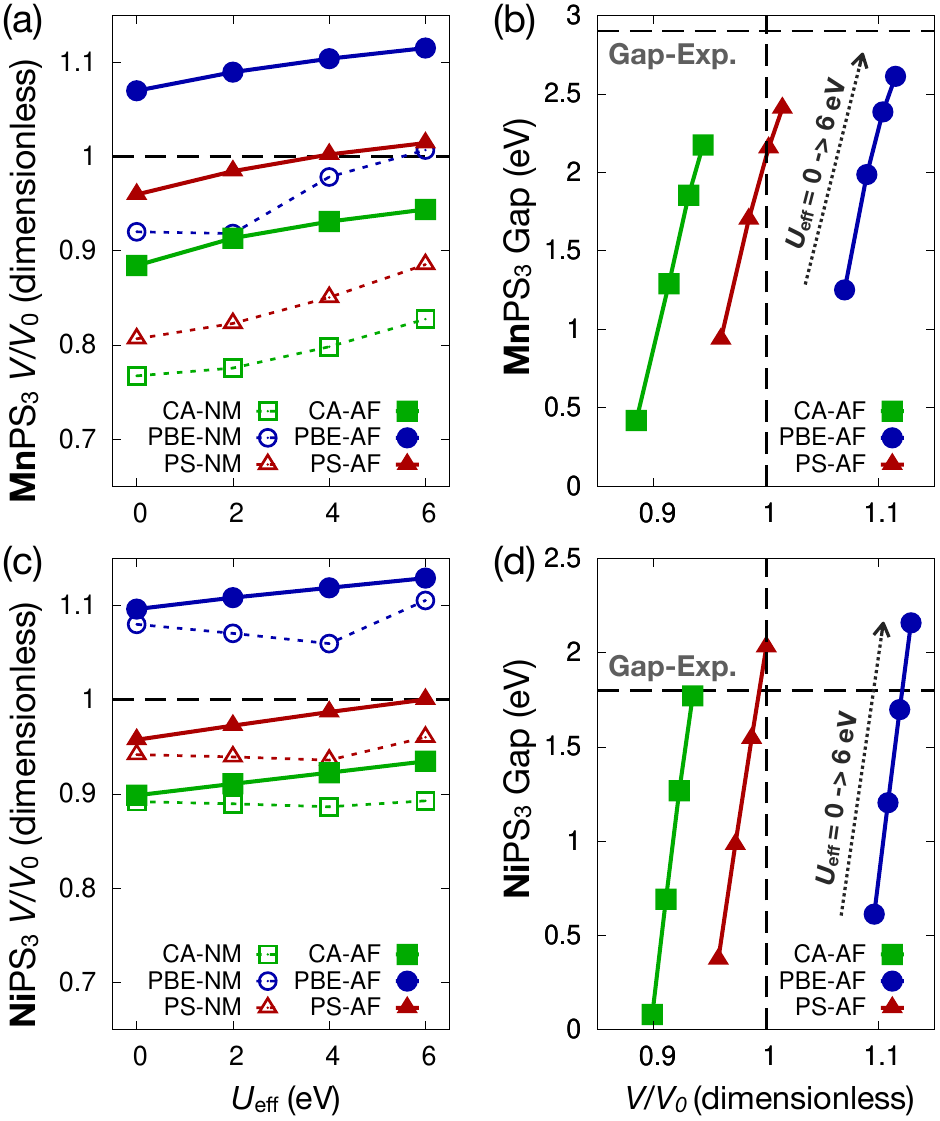}
\caption{
Calculated unit cell volumes and band gap sizes of (a,b) MnPS$_3$ and (c,d) NiPS$_3$ under different choices of exchange-correlation (XC) functionals, the presence of antiferromagnetic order, and $U_{\rm eff}$ values at the ambient pressure condition ({\it i.e.} optimizing cell volume without any volume/shape constraint). (a,c) show how the cell volume depends on the choice of XC functionals, the presence of antiferromagnetic order, and $U_{\rm eff}$ values for each compound, and (b,d) show the size of gap as a function of cell volume with respect to the choice of XC functionals. Therein 4 data points for each XC functional, from bottom to top, represent results with $U_{\rm eff}$ = 0, 2, 4, and 6 eV respectively. In the legend PS, NM, AF denote PBEsol, nonmagnetic, and antiferromagnetic orders respectively. N\'{e}el- and zigzag-type AF order were employed for MnPS$_3$ and NiPS$_3$, respectively. $V_0$ denotes experimental cell volume for each compound\cite{BREC1986}, and horizontal black dashed lines in (b,d) show experimentally measured gap sizes. 
 }
 \label{fig:xc_U}
 \end{figure}

Fig. \ref{fig:xc_U} shows how MnPS$_3$ and NiPS$_3$ behave under the choice of different exchange-correlation functionals and the value of $U_{\rm eff}$,
where experimental cell volumes and gap sizes are from Ref. \onlinecite{BREC1986,Grasso1991,Soyeun2018}. Fig. \ref{fig:xc_U}(a) and (c) show how the cell volume depends on the choice of exchange-correlation (XC) functionals, the presence of antiferromagnetic (AF) order, and the $U_{\rm eff}$ values for each compound. We notice that {\it i)} the absence of AF order, which prevents the formation of high-spin configurations in both compounds, yields significantly underestimated cell volumes in all cases. Such behavior is more evident in MnPS$_3$, where the absence of magnetism leads to the low-spin configuration that favors intermetallic bonding. {\it ii)} the PBEsol XC functional gives better agreement with experimental volume than CA or PBE, and using $U_{\rm eff}$ = 4 $\sim$ 6 eV in the PBEsol+$U_{\rm eff}$ setup produces the best fit. The same conclusion can be also made from the gap-volume dependence shown in Fig. \ref{fig:xc_U}(b) and (d), where the use of PBEsol+$U_{\rm eff}$ (4 eV) yields the best fit of gap size and cell volume for both compounds. Overall, employing PBEsol+$U_{\rm eff}$ (4 eV) does seem reasonable for studying pressurized MnPS$_3$ and NiPS$_3$.

\begin{figure}
\includegraphics[width=0.48\textwidth]{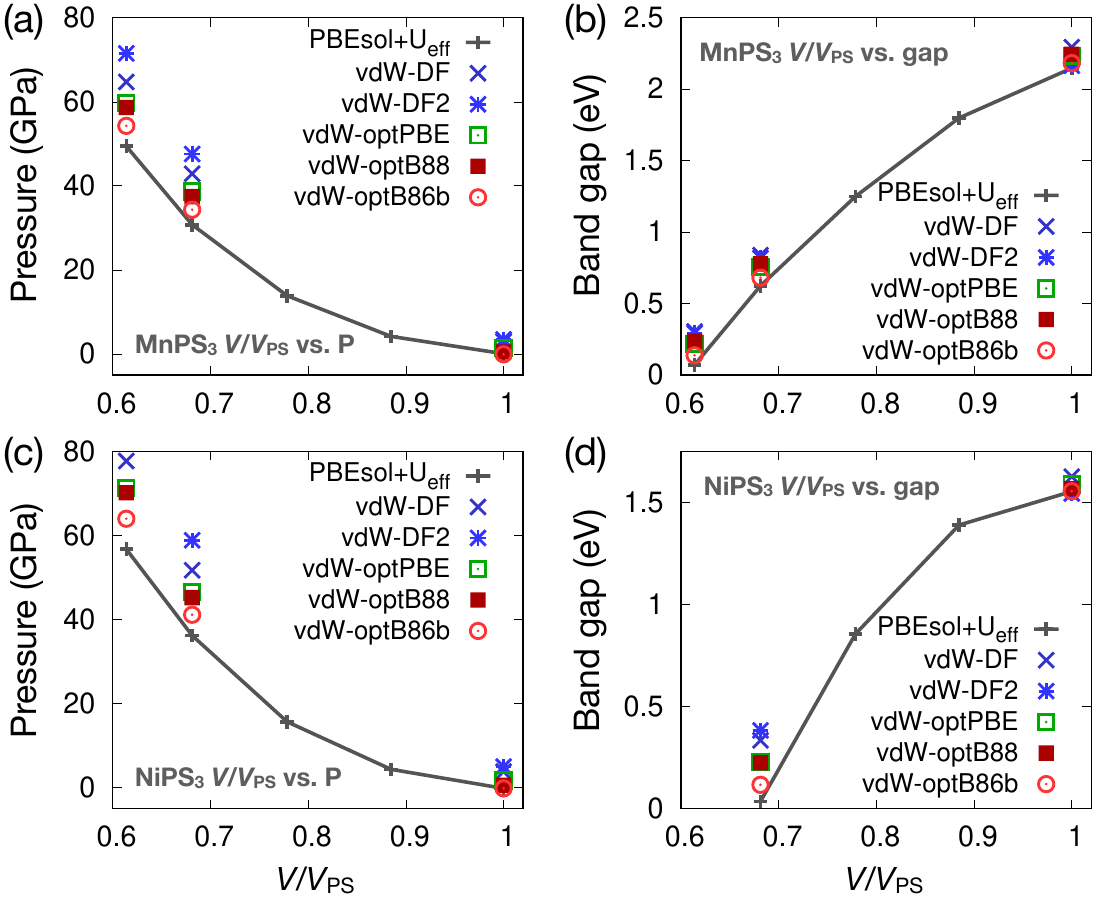}
\caption{Dependence of crystal and electronic structures on five van der Waals (vdW) functionals (see the text) for (a,b) MnPS$_3$ and (c,d) NiPS$_3$. Colored symbols represent vdW functional results, and PBEsol+$U_{\rm eff}$ (4 eV) results (dark gray curves) are shown as a reference. In panel (a) and (c), the calculated pressure $P$ is shown as a function of dimensionless volume ($V/V_{\rm PS}$, where $V_{\rm PS}$ is the computed ambient-condition volume with PBEsol+$U_{\rm eff}$ = 4 eV). In panel (b) and (d), band gap is shown as a function of $V/V_{\rm PS}$.  
 }
 \label{fig:xc_vdw}
 \end{figure}
 
 \subsection{van der Waals functional dependence}

Fig.~\ref{fig:xc_vdw} presents how the estimated pressure and the size of Kohn-Sham energy gap at a given volume depend on the choice of van der Waals (vdW) functionals in MnPS$_3$ and NiPS$_3$. The range of unit cell volume is chosen to be $0.6 V_{\rm PS} \leq V \leq 1.0 V_{\rm PS}$, where $V_{\rm PS}$ is the ambient-pressure cell volume optimized with PBEsol+$U_{\rm eff}$ (4 eV). We employed 5 different vdW functionals implemented in {\sc vasp}; vdW-DF\cite{vdW-DF}, vdW-DF2\cite{vdW-DF2}, optPBE, optB88, and optB86b\cite{vdW-opt}. For both compounds, different functionals tend to give similar results, while NiPS$_3$ shows more noticeable functional dependence compared to MnPS$_3$. It can be speculated that, in NiPS$_3$ with an open-shell Ni $e_{\rm g}$ orbitals, the unquenched orbital degree of freedom makes the system a bit more sensitive to the treatment of correlations. Nevertheless the qualitative features we address in this manuscript, the pressure-driven insulator-to-metal transitions and their orbital dependence, remain basically the same as shown in Fig. \ref{fig:xc_vdw}(b) and (d). 

Specifically we notice that vdW functionals, except vdW-DF and vdW-DF2 functionals, show reasonable agreements with PBEsol results. vdW-DF and vdW-DF2 functionals tends to prefer larger volume ({\it i.e.} larger $P$ estimated at the same volume compared to other functionals). This is because these functionals favor larger interlayer distances in the high-pressure regime than conventional XC functionals. Overall, even though use of different vdW functional induces some quantitative differences, it does not seem to change our main conclusions in this work.

We comment that, since vdW functionals favor larger cell volume, the inclusion of them should enhance the magnitude of critical pressures for structural/electronic transitions in both compounds, which is actually making the discrepancy between the theoretical prediction and experimental observation slightly worse in MnPS$_3$. Hence we argue that PBEsol+$U_{\rm eff}$ can be a more reasonable choice in this pressurized setup where the direct orbital overlap between layers becomes significant.

\section{Comparison with DFT+$U_{\rm eff}$ and DMFT}

%

\renewcommand*{\arraystretch}{1.4}
\begin{table*}[htb!]
\centering
\begin{tabular}{lrrrrrlrrrrr}  \hline\hline
&                                            & \multicolumn{4}{c}{MnPS$_3$} &&& \multicolumn{4}{c}{NiPS$_3$} \\ \hline
\multicolumn{2}{l}{$V/V_{\rm PS}$}                 	
					    & \multicolumn{2}{c}{1.00} & \multicolumn{2}{c}{0.56} &&
					    & \multicolumn{2}{c}{1.00} & \multicolumn{2}{c}{0.56} \\ [-3pt]
\multicolumn{2}{l}{$P_{\rm DFT}$ (GPa)}        
					    & \multicolumn{2}{c}{0.0} & \multicolumn{2}{c}{49.8} &&
					    & \multicolumn{2}{c}{0.0} & \multicolumn{2}{c}{88.0} \\ [3pt]
\multicolumn{2}{l}{\bf a}                 
					    & \multicolumn{2}{c}{6.025} & \multicolumn{2}{c}{4.916} &&
					    & \multicolumn{2}{c}{5.761} & \multicolumn{2}{c}{5.173} \\ [-3pt]
\multicolumn{2}{l}{\bf b}                 
					    & \multicolumn{2}{c}{10.436} & \multicolumn{2}{c}{8.845} &&
					    & \multicolumn{2}{c}{9.977} & \multicolumn{2}{c}{8.531} \\ [-3pt]
\multicolumn{2}{l}{{\bf c} (\AA)}        
					    & \multicolumn{2}{c}{6.870} & \multicolumn{2}{c}{5.542} &&
					    & \multicolumn{2}{c}{6.736} & \multicolumn{2}{c}{4.938} \\ [-3pt]
\multicolumn{2}{l}{${\beta}$ (degree)} 
					    & \multicolumn{2}{c}{106.67} & \multicolumn{2}{c}{108.72} &&
                 			    & \multicolumn{2}{c}{106.64} & \multicolumn{2}{c}{110.17} \\ \hline
&& DFT+$U_{\rm eff}$ & eDMFT &DFT+$U_{\rm eff}$ & eDMFT &&& DFT+$U_{\rm eff}$ & eDMFT &DFT+$U_{\rm eff}$ & eDMFT  \\[3pt]
Mn   ($4g$)  & $y$ & 0.3327 & 0.3326 & 0.3589 & 0.3585 &
~~~~~~~~~~
Ni    ($4g$)  & $y$ & 0.3333 & 0.3329 & 0.3339 & 0.3344 \\[3pt]
P     ($4i$)  & $x$ & 0.0550 & 0.0552 &-0.0376 & -0.0256 &
~~~~~~~~~~
P     ($4i$)  & $x$ & 0.0570 & 0.0573 & 0.0803 & 0.0790 \\[-3pt]
	 	  & $z$ & 0.1674 & 0.1683 & 0.1719 & 0.1770 &
		  & $z$ & 0.1677 & 0.1687 & 0.2223 & 0.2199 \\[3pt]
S1   ($4i$)  & $x$ & 0.7551 & 0.7431 & 0.6707 & 0.6831 &
~~~~~~~~~~
S1   ($4i$)  & $x$ & 0.7346 & 0.7378 & 0.7299 & 0.7324 \\[-3pt]
	 	  & $z$ & 0.2474 & 0.2512 & 0.3409 & 0.3539 &
		  & $z$ & 0.2371 & 0.2420 & 0.2948 & 0.2995 \\[3pt]
S2   ($8j$)  & $x$ & 0.2441 & 0.2448 & 0.2426 & 0.2463 &
~~~~~~~~~~
S2   ($8j$)  & $x$ & 0.2523 & 0.2467 & 0.2990 & 0.2988 \\[-3pt]
	 	  & $y$ & 0.1628 & 0.1625 & 0.1858 & 0.1855 &
		  & $y$ & 0.1727 & 0.1719 & 0.1930 & 0.1923 \\[-3pt]
	 	  & $z$ & 0.2485 & 0.2525 & 0.2617 & 0.2781 &
		  & $z$ & 0.2366 & 0.2422 & 0.2909 & 0.2955 \\ \hline\hline
\end{tabular}
\caption{Optimized lattice parameters of MnPS$_3$ and NiPS$_3$ from DFT+$U_{\rm eff}$ and eDMFT results, both at ambient and high-pressure regimes. Ambient and high-pressure results represent Mott-insulating and weakly correlated metallic phases, respectively, for both compounds. PBEsol+$U_{\rm eff}$ = 4 eV was adopted for DFT+$U_{\rm eff}$. Cell parameters ({\bf a}, {\bf b}, {\bf c}, and $\beta$) optimized in DFT+$U_{\rm eff}$ calculations were employed in eDMFT ones. Nonzero components of Wyckoff positions of the $C2/m$ space group are shown. All eDMFT calculations were done at $T$ = 232K except the high-pressure ($V = 0.56V_{\rm PS}$) MnPS$_3$ one, where $T$ = 580K was used for computational issues. $V_{\rm PS}$ denotes the ambient pressure cell volume for both compounds, obtained with PBEsol+$U_{\rm eff}$ = 4 eV 
}
\label{tabS:str}
\end{table*}

Table \ref{tabS:str} presents the comparison between PBEsol+$U_{\rm eff}$- and eDMFT-optimized atomic coordinates of MnPS$_3$ and NiPS$_3$, both at ambient and high-pressure regimes. Here ambient and high-pressure results represent Mott-insulating and weakly correlated metallic phases, respectively, for both compounds. In eDMFT calculations, as commented in the manuscript, optimized cell parameters {\bf a}, {\bf b}, {\bf c}, and monoclinic angle $\beta$ from PBEsol+$U_{\rm eff}$ were employed. This is due to the absence of stress tensor formalism implemented in any of full-potential linearized augmented plane wave codes, and in DFT+DMFT formalisms as well. Under this constraint, eDMFT-optimized atomic coordinates show very similar results with PBEsol+$U_{\rm eff}$ ones despite different magnetization conditions; paramagnetic order for eDMFT, and antiferromagnetism (N\'{e}el order for MnPS$_3$, zigzag order for NiPS$_3$) in PBEsol+$U_{\rm eff}$. 

However, the validity of employing DFT+$U_{\rm eff}$ with a magnetic order in optimizing crystal structures of paramagnetic systems, especially the cell parameters, may need to be checked. This is because, like in MnPS$_3$ as shown in the manuscript, some structural phase transitions are strongly coupled to elastic deformations of the unit cell. One may even suspect that the discrepancy between the predicted and experimentally reported\cite{WangJACS2016} critical pressures of the structural transition in MnPS$_3$ might originate from the use of magnetic DFT+$U_{\rm eff}$ in optimizing the unit cell size and shape. 

To resolve the issue mentioned above, energy landscapes from DFT+$U_{\rm eff}$ and eDMFT in the cell parameter space need to be compared with each other. Even though full structural relaxations may not be possible within eDMFT formalism, several trials to compare DFT+$U_{\rm eff}$- and DMFT-optimized structures were performed, where DMFT calculations were done in the paramagnetic configuration. 

 \subsection{NiPS$_3$}
 
 \begin{figure}
  \centering
  \includegraphics[width=0.5\textwidth]{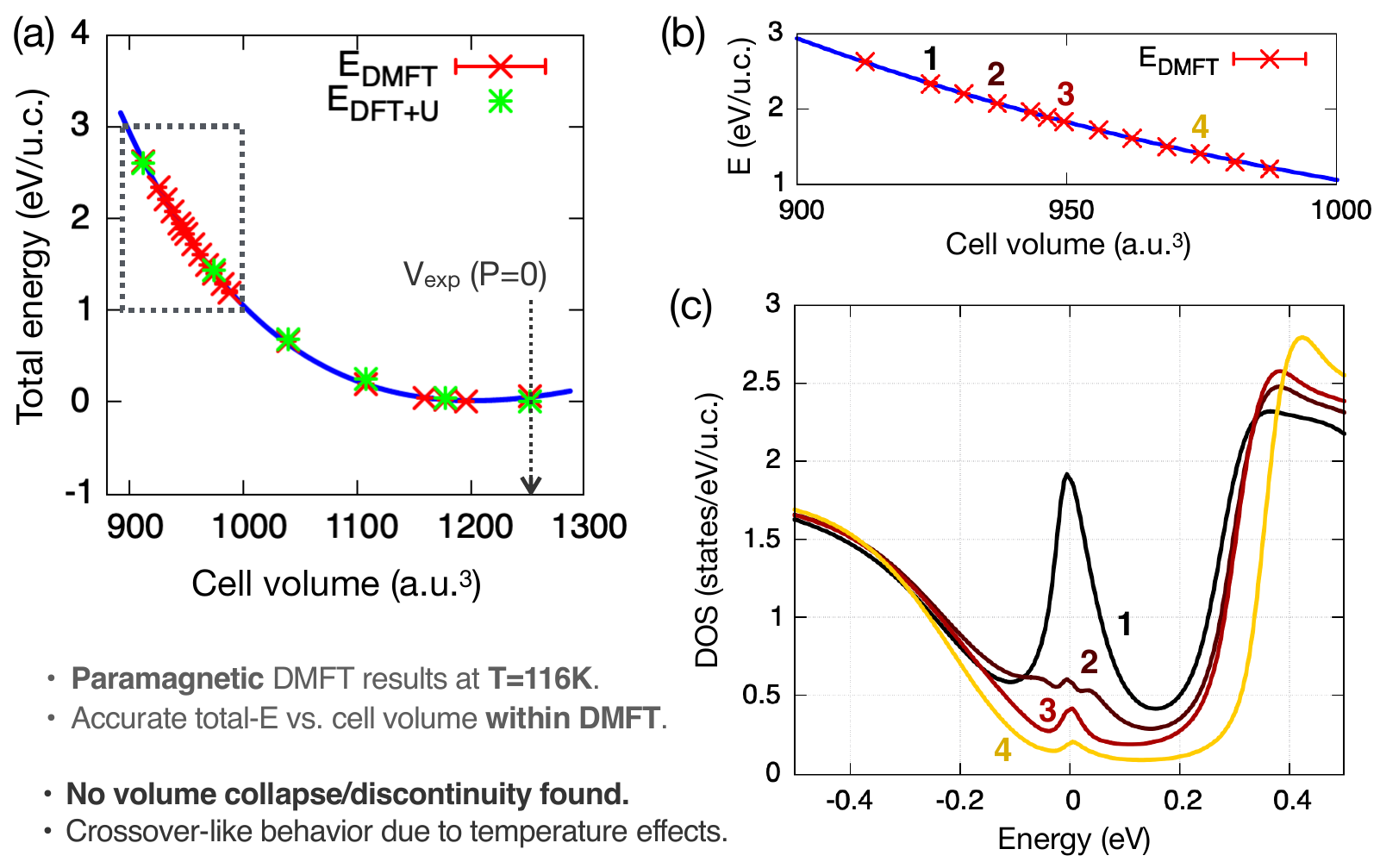}
  \caption{
	(a) Calculated energy versus volume plot for NiPS$_3$. Bright green and red symbols represent data points from PBEsol+$U_{\rm eff}$ and eDMFT results respectively. Sizes of error bars in eDMFT points ($\lesssim$ 5 meV) are smaller than the symbol size. Blue curve is from the Birch-Murnaghan fit of eDMFT energies. Note that eDMFT result predicts slightly smaller ambient-pressure volume compared to experimental value. (b) A magnification of the eDMFT energy-volume data close to the insulator-to-metal transition, where the area of magnification indicated as a gray dashed box in (a). (c) Real-frequency spectral weights at different volumes close to the insulator-to-metal transition. Colored number for each curve indicates at which volume the spectral function was taken (see (b)). Calculated pressures at point 1 and 2 from the Birch-Murnaghan fit are 24.3 and 22.2 GPa, respectively. 
  }
  \label{fig:NPS_EvsV}
\end{figure}

According to our result presented in the manuscript, NiPS$_3$ does not have a noticeable structural change at the MIT pressure of $P$ $\simeq$ 30 GPa, which is somewhat unusual. Hence, for a closer look on the structure-free MIT point, we computed a total energy curve versus cell volume for paramagnetic (PM) eDMFT near the MIT. For more accurate results, the rotationally invariant form of the Coulomb interaction (spin-flip and pair-hopping included) was employed at a lower temperature of $T$ = 116 K. PM MIT is usually known to accompany a sudden volume change (as reported in Ref.~35 in the manuscript, for example), which should be captured as a discontinuity in the energy-volume curve at the MIT point. Figure \ref{fig:NPS_EvsV} shows a summary of the results; note that all the data points were obtained from calculations started from scratch to capture both the metallic and insulating phases (with optimized internal parameters). It can be seen that both the PBEsol+$U_{\rm eff}$ and DFT+DMFT data points remarkably collapse onto a single Birch-Murnaghan energy-volume curve (Fig.~\ref{fig:NPS_EvsV}(a)), and that no discontinuity can be noticed near the MIT point (Fig.~\ref{fig:NPS_EvsV}(b) and (c)). Note that the use of the additional Coulomb interaction with the spin-flip and pair-hopping terms (hereafter denoted as `beyond-Ising' terms) lowered the MIT critical pressure from 31 to 24 GPa. The crossover-like behavior can be attributed to the temperature effect, but due to the computational cost issue the temperature could not be lowered below $T$ = 116 K. 

 
 \subsection{MnPS$_3$}

\begin{figure}
\includegraphics[width=0.45\textwidth]{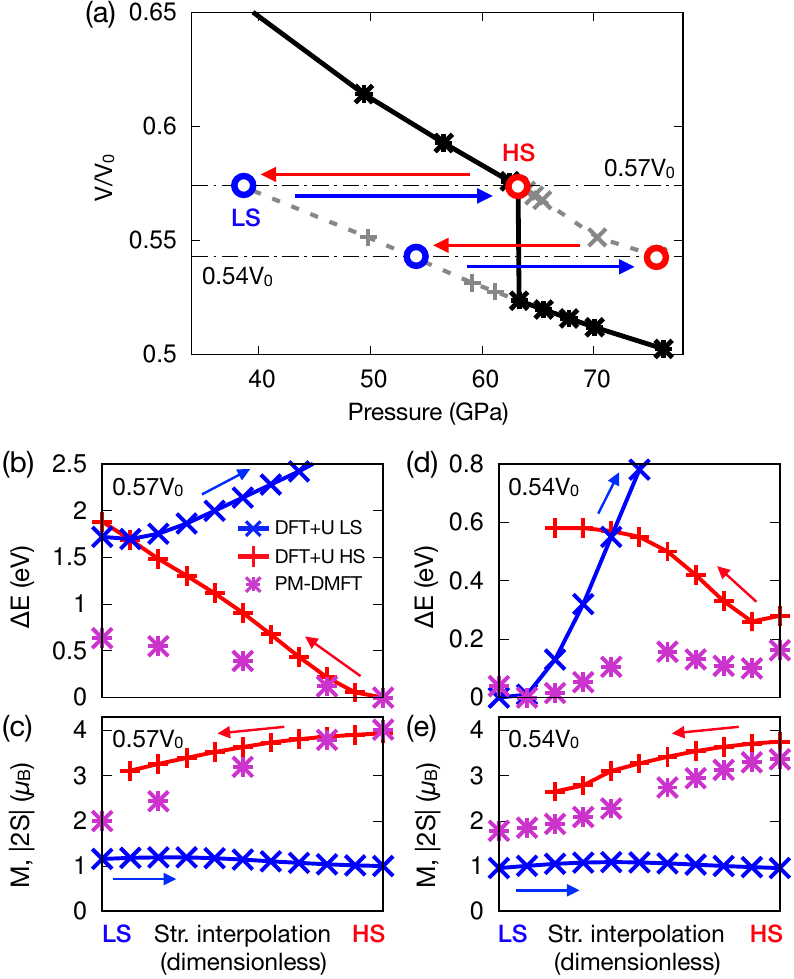}
\caption{\label{fig:interp}(a) A magnified view of the MnPS$_3$ volume vs. pressure plot, where the black solid and gray dashed lines are curves for ground and metastable states respectively. Two dash-dotted lines are at $V$ = 0.57 and 0.54 $V_0$ on which structural interpolations between the high-spin honeycomb and low-spin dimerized structures are made. (b,c) Total energy differences (b) and size of Mn moments (c) as a function of structural interpolation at $V$ = 0.57 $V_0$, where red (blue) curve depicting total energy and Mn magnetization starting from high-spin (low-spin) structure and approaching to the low-spin (high-spin) side, and purple symbols depicting same results from paramagnetic eDMFT calculations at $T$ = 580K. (d,e) Same plots at $V$ = 0.54 $V_0$. Size of QMC error bars for eDMFT results are 4 meV at most, smaller then the symbol size.  
}
\end{figure}

As marked in Fig.~1(a) in the main text, the value of critical pressure predicted by PBEsol+$U_{\rm eff}$ calculations is twice bigger than the one reported in Ref. \onlinecite{WangJACS2016} (63 vs. 30 GPa). For a better understanding of this discrepancy, we perform calculations with interpolating structures between the honeycomb high-spin and the dimerized low-spin structures as shown in Fig.~\ref{fig:interp}. Because the computation of stress tensor is not yet available in the current eDMFT formalism, two constant-volume cuts at $V$ = 0.57 and 0.54 $V_0$ are taken for a total energy comparison as shown in Fig.~\ref{fig:interp}(a). For PBEsol+$U_{\rm eff}$ results, high-spin and low-spin states are first converged in their honeycomb and dimerized structures respectively, and then the crystal structures are slowly distorted towards the other side while maintaining the local minima spin states. 

Fig.~\ref{fig:interp}(b) and (c) are relative total energies and size of spin moments from the high-, low-spin PBEsol+$U_{\rm eff}$, and paramagnetic eDMFT calculations at $V$ = 0.57 $V_0$.  A remarkable feature is, while the energy difference between the high- and low-spin ground states is 1.72 eV in PBEsol+$U_{\rm eff}$, it is 0.58 eV in eDMFT, which is almost one third of the PBEsol+$U_{\rm eff}$ value. Furthermore, while the high- and low-spin local minima states remain (meta)stable even after the structural changes, as shown in Fig.~\ref{fig:interp}(c), in eDMFT we have a spin-state crossover as the structure evolve from one limit to another. This features persist at $V$ = 0.54 $V_0$ (Fig.~\ref{fig:interp}(d,e)), where the height of the energy barrier from the high-spin to the low-spin state (0.3 eV) is substantially suppressed (60 meV) with the same spin-state crossover. These observations show that, the dynamical fluctuation effect inherent in eDMFT causes mixing between different spin configurations, hence introducing the crossover behavior shown in Fig.~\ref{fig:interp}(c) and (e) and suppressing the energy differences. This observation is consistent with a previous DFT+DMFT study on a spin-state-crossover molecule\cite{JChen2015}. Note that, our analysis here does not explicitly predict that the low-spin state is stabilized at lower pressure in eDMFT results shown in Fig.~1(a) in the main text. However, since the energy difference between different states is reduced to a fraction compared to PBEsol+$U_{\rm eff}$ results, the value of critical pressure might be reduced after the lattice free energy contribution and the zero-point fluctuation ignored in this work are included.
%


\begin{table*}[htp]
\footnotesize
\begin{center}
\label{tab:strs}
\begin{tabular}{lcccc}
& \multicolumn{2}{c}{Dimerized} & \multicolumn{2}{c}{Non-dimer} \\ \hline\hline
$V$ & \multicolumn{2}{c}{0.55$V_{\rm PS}$} & \multicolumn{2}{c}{0.61$V_{\rm PS}$} \\
$P_{\rm PS}$ & \multicolumn{2}{c}{49.8 GPa} & \multicolumn{2}{c}{49.4 GPa} \\
${\bf a}/{\bf a}_0$ & \multicolumn{2}{c}{0.8159} & \multicolumn{2}{c}{0.8619} \\
${\bf b}/{\bf b}_0$ & \multicolumn{2}{c}{0.8475} & \multicolumn{2}{c}{0.8653} \\
${\bf c}/{\bf c}_0$ & \multicolumn{2}{c}{0.8066} & \multicolumn{2}{c}{0.8263} \\
$\beta$                & \multicolumn{2}{c}{108.72$^\circ$} & \multicolumn{2}{c}{107.32$^\circ$} \\ \hline
Coulomb &Ising & Beyond-Ising & ~~~~~~~~Ising & Beyond-Ising \\ 
Mn (4g) & (0.0000, 0.1415, 0.0000) & (0.0000, 0.1407, 0.0000) & ~~~~~~~~(0.0000, 0.1664, 0.0000) &(0.0000, 0.1661, 0.0000) \\
P (4i)     & (0.5256, 0.0000, 0.8230) & (0.5278, 0.0000, 0.8238) & ~~~~~~~~(0.4358, 0.0000, 0.8064) &(0.4358, 0.0000, 0.8063) \\
S1 (4i)   & (0.8170, 0.0000, 0.6461) & (0.8176, 0.0000, 0.6451) & ~~~~~~~~(0.2598, 0.0000, 0.6964) &(0.2603, 0.0000, 0.6961) \\
S2 (8j)   & (0.7538, 0.3145, 0.7219) & (0.2542, 0.3142, 0.7227) & ~~~~~~~~(0.2231, 0.3205, 0.6980) &(0.2228, 0.3201, 0.6984) \\ \hline
\end{tabular}
\caption{
Optimized crystal structures of MnPS$_3$ under different volume constraints ($V = 0.55V_{\rm PS}$ and $0.61V_{\rm PS}$, where $V_{\rm PS}$ and $P_{\rm PS}$ are optimized ambient-pressure volume and exerted pressure at a fixed volume from PBEsol+$U_{\rm eff}$ (4 eV) results. Equilibrium lattice parameters from PBEsol+$U_{\rm eff}$ calculations are given as $({\bf a}_0, {\bf b}_0, {\bf c}_0, \beta_0) = (6.025{\rm \AA}, 10.436{\rm \AA}, 6.870{\rm \AA}, 106.67^\circ)$. Internal coordinates optimized from eDMFT calculations, with different choices of Coulomb interactions (Ising vs. beyond-Ising), are listed below. 
}
\end{center}
\end{table*}%

We comment that, due to the increased computational cost of the CT-HYB impurity solver by the enhanced intermetallic hybridization in the pressurized setup, all eDMFT data points presented in Fig.~\ref{fig:interp} were obtained at $T$ = 580 K. Even without the beyond-Ising Coulomb terms, which significantly increases sign problems in the impurity solver stage, the temperature could not be lowered due to the computational cost issue. Therefore we could not check all the results with employing the beyond-Ising Coulomb terms. As a partial check, we have done two eDMFT calculations for MnPS$_3$, employing two lattice parameter sets ($\bf a$, $\bf b$, $\bf c$, and the monoclinic angle $\beta$) corresponding to non-dimerized metallic and dimerized Mott-insulating states around $50\,$GPa. Initial internal coordinates, optimized within eDMFT afterwards, were adopted from the ambient pressure structure. The purpose of this comparison is to see whether the choice of different Coulomb interactions yields noticeable difference. Table II summarizes the optimized structures with Ising and beyond-Ising Coulomb interactions, which shows negligible difference with respect to each other. Hence we are safe to use Ising form in this case. 


\begin{figure}
\includegraphics[width=0.45\textwidth]{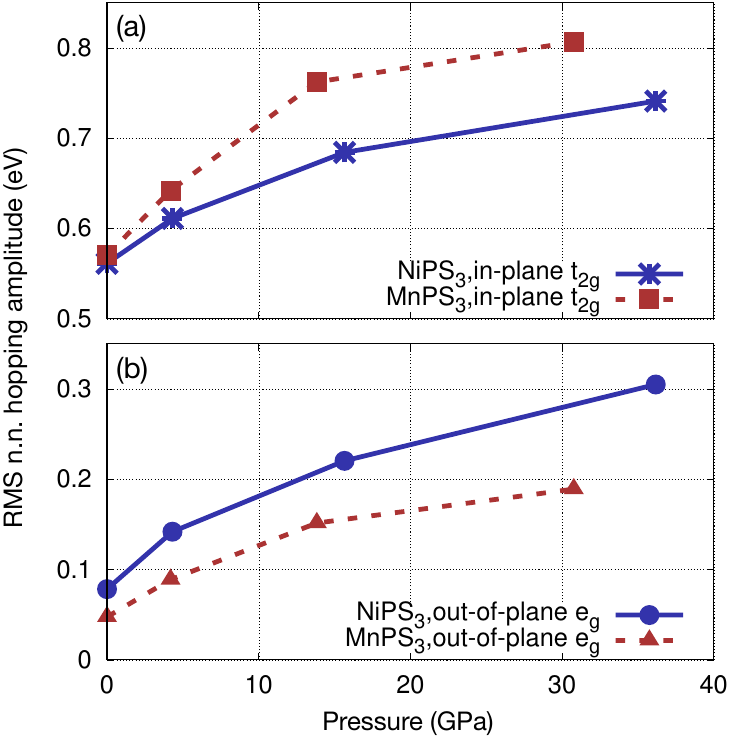}
\caption{\label{fig:hop}Root mean square (RMS) of nearest-neighbor (n.n.) Wannier hopping amplitudes for Ni and Mn $d$-orbitals as a function of pressure, where the RMS of (a) in-plane $t_{\rm 2g}$-$t_{\rm 2g}$ and (b) out-of-plane $e_{\rm g}$-$e_{\rm g}$ are shown.}
\end{figure}

\section{In-plane and out-of-plane hopping amplitudes}
Fig.~\ref{fig:hop}(a) and (b) show root mean square (RMS) amplitudes of nearest-neighbor (n.n.) in-plane $t_{\rm 2g}$ and out-of-plane $e_{\rm g}$ hopping integrals, respectively, for MnPS$_3$ and NiPS$_3$. The $d$-orbital hopping integrals were computed using {\sc wannier90} package\cite{Wannier90}, employing optimized crystal structures in the presence of external pressure, without including $U_{\rm eff}$ and magnetism. 

As shown in Fig.~1(a) and (b) in the main text, the in-plane lattice parameters (with respect to their ambient pressure value) $a/a_0$ and $b/b_0$ for MnPS$_3$ around 30 GPa are smaller by $\sim$ 2\% compared to those of NiPS$_3$, while the out-of-plane $c/c_0$ of MnPS$_3$ is larger than that of NiPS$_3$. In accordance with the tendency of lattice parameter changes, the enhancement of RMS in-plane $t_{\rm 2g}$ hopping integrals is more pronounced in MnPS$_3$, which drives the formation of in-plane Mn dimer formation after the transition to the low-spin state with the open $t_{\rm 2g}$ shell. Note that, the in-plane $t_{\rm 2g}$ hopping integrals for Ni is also enhanced as the pressure is increased, but its effect is not significant due to the closed $t_{\rm 2g}$ shell in the Ni $d^8$ configuration. Similarly, the enhanced out-of-plane kinetic energy between the $e_{\rm g}$ orbitals in NiPS$_3$, depicted in Fig.~\ref{fig:hop}(b), induces more pronounced reduction of the $c$ parameter in NiPS$_3$ compared to that of MnPS$_3$. It should be mentioned that, while the out-of-plane $e_{\rm g}$ hopping terms are also enhanced in MnPS$_3$, their role in structural response to pressure is less significant both in the low- and high-pressure regimes; in the low-pressure regime the hybridization between the $d^5$ high-spin Mn ion and anions is weak, and so is the electron-lattice coupling, while in the high-pressure regime the high-spin Mn has empty $e_{\rm g}$ shell. 

\begin{figure*}
\includegraphics[width=0.75\textwidth]{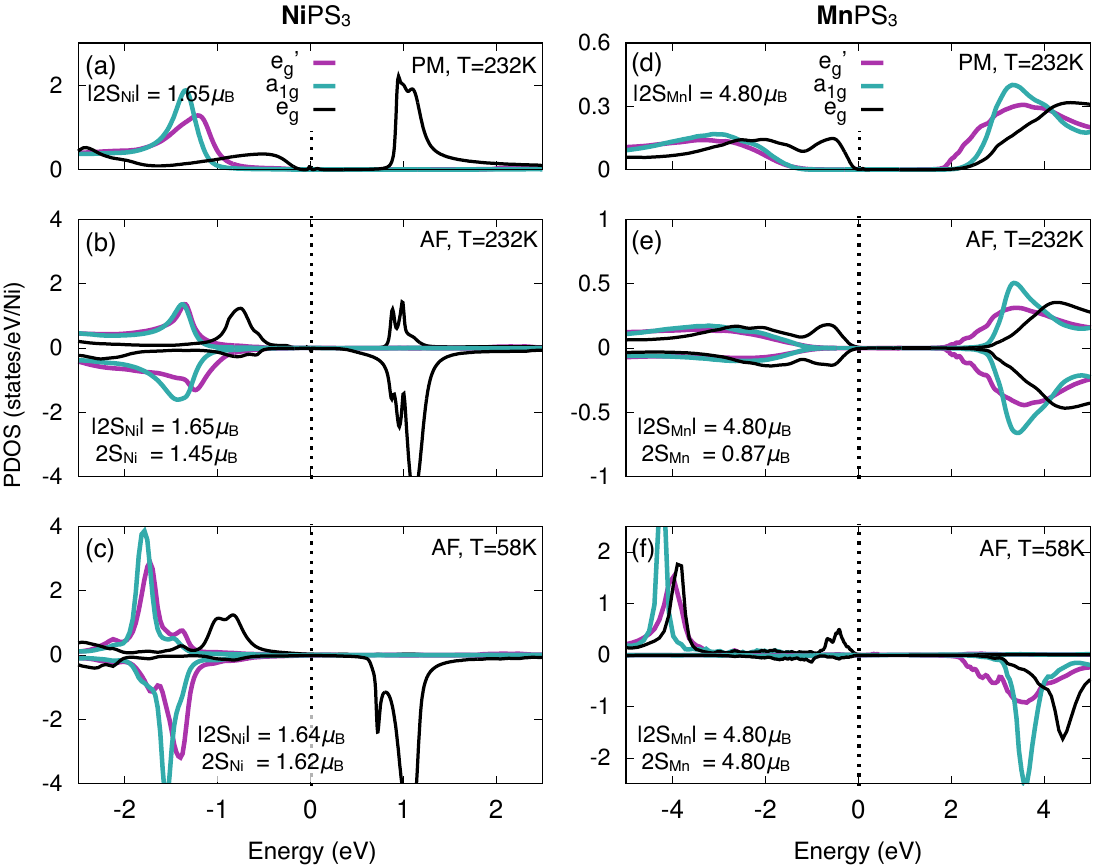}
\caption{\label{fig:mag1}(a-c) PDOS of NiPS$_3$ and (d-f) MnPS$_3$ with a N\'{e}el type antiferromagnetic (AF) order (b,c,e,f) in comparison with paramagnetic (PM) PDOS (a,d). The second and third rows show AF PDOS with $T$ = 232K and 58K, respectively. 
}
\end{figure*}

\begin{figure*}
\includegraphics[width=0.9\textwidth]{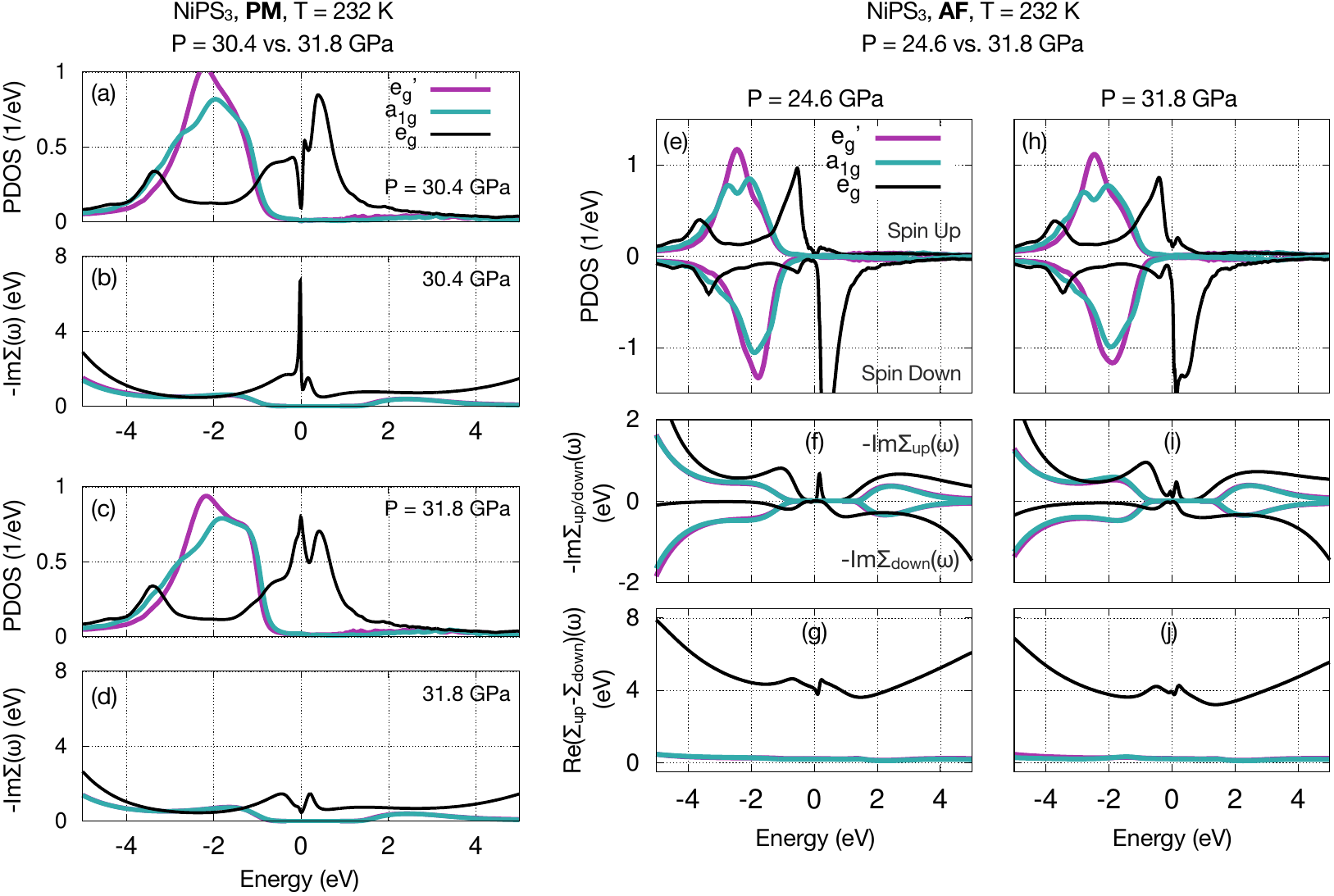}
\caption{\label{fig:mag2}PDOS and electron self-energies $\Sigma(\omega)$ of (a-d) PM and (e-j) AF NiPS$_3$ at $T$ = 232K. The imaginary part of PM self-energies in (b) and (d) show a clear $e_{\rm g}$ peak and its absence in the Mott-insulating and metallic phases, respectively. (e) and (h) show spin- and orbital-projected PDOS at P = 24.6 and 31.8 GPa, respectively (spin up and down PDOS plotted in positive and negative values respectively). Imaginary part of $\Sigma(\omega)$, plotted in (f) and (i), show the absence of peak near the Fermi level. (e) and (h) show the difference between the real part of spin up and down self-energies $\Re\left(\Sigma_{\rm up} - \Sigma_{\rm down}\right)(\omega)$, which is the on-site exchange splitting opening the gap in the magnetic phases. Note that, the MIT point is not clear due to the smearing of spin up and down $e_{\rm g}$ bands in the PDOS.
}
\end{figure*}

\section{Electronic structures with a N\'{e}el order}
\subsection{Ambient pressure cases for both compounds}
Fig.~\ref{fig:mag1} shows the spectral functions of NiPS$_3$ (a-c) and MnPS$_3$ (d-f) with a N\'{e}el-type antiferromagnetic (AF) order in comparison with the paramagnetic (PM) phases. In both compounds, the presence of magnetism does not alter the size of charge gap, consistent with the Mott character of the insulating phases in both compounds. As the temperature is lowered and magnetism sets in, the broadening of spectral functions due to the imaginary part of self-energy is weakened. Indeed, DFT+$U_{\rm eff}$ PDOS with the magnetic order shows very similar qualitative features with DMFT PDOS at $T$=58K (not shown). 

Note that, the use of the Ising-type Coulomb repulsion gives rise to the stabilization of magnetic order well above the N\'{e}el temperatures of both compounds, $T_N$ = 154 and 78 K for NiPS$_3$ and MnPS$_3$, respectively\cite{BREC1986,Wildes2006,Wildes2015}, as reported in previous DFT+DMFT studies\cite{Hab2018PRL}. It is argued that a larger in-plane kinetic energy scale originating from the $e_{\rm g}$ orbital in Ni yields the higher $T_N$ in NiPS$_3$ compared to MnPS$_3$\cite{Soyeun2018}. Despite the overestimated $T_N$, such tendency can be noticed in our results by comparing Fig.~\ref{fig:mag1}(b) and (e). While the size of the Ni magnetization in NiPS$_3$ (2$S_{\rm Ni}$ = 1.45 $\mu_{\rm B}$) is almost saturated to the value of PM moment ($\vert 2S_{\rm Ni} \vert$ = 1.65 $\mu_{\rm B}$), even at $T$ = 232K, the Mn magnetization in MnPS$_3$ at the same $T$ is 2$S_{\rm Mn}$ = 0.87 $\mu_{\rm B}$, just a fraction of the $S$ = 5/2 moment size (4.80 $\mu_{\rm B}$) of the high-spin Mn. As $T$ is lowered to 58K, magnetizations in both compounds saturate to the local moment size, as shown in Fig.~\ref{fig:mag1}(c) and (f).

\subsection{Near the MIT critical pressure}

\subsubsection{MnPS$_3$}
\begin{figure}
\includegraphics[width=0.4\textwidth]{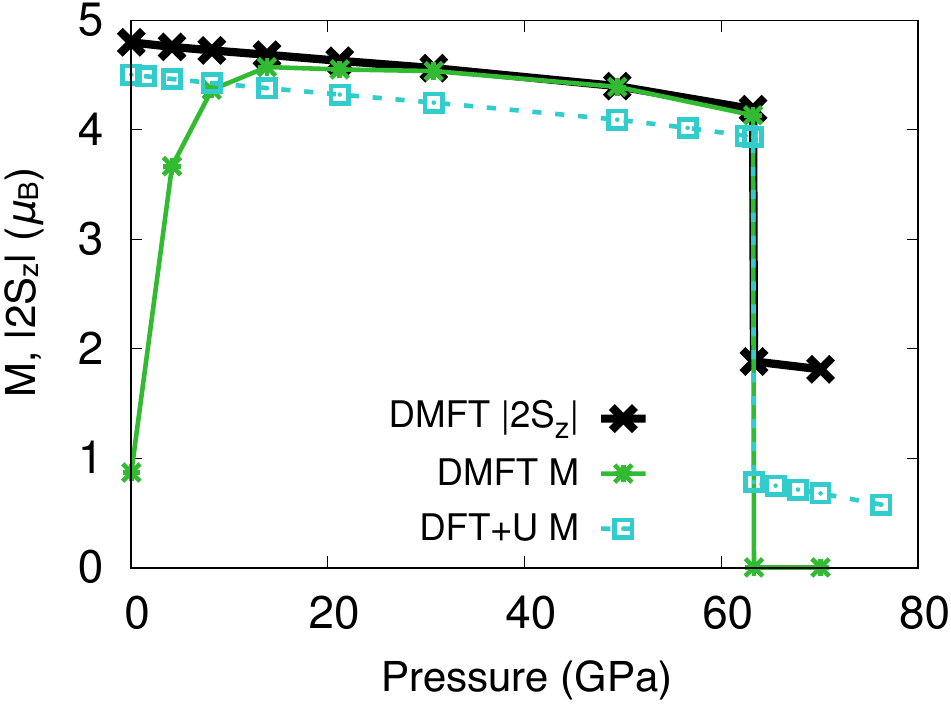}
\caption{\label{fig:magMn}Evolution of sizes of DFT+$U_{\rm eff}$ and eDMFT local moments $\vert 2S_{z} \vert$ and Mn magnetization $M$ (per Mn) in MnPS$_3$ as a function of pressure. eDMFT results are obtained at $T$ = 232 K with the N\'{e}el order. 
}
\end{figure}

As discussed in the main text, the pressure-induced MIT in MnPS$_3$ in the paramagnetic phase is a discontinuous transition accompanied by a spin-state transition and isosymmetric structural distortion with a volume collapse. Such discontinuous character does not change in the presence of magnetism, as shown in Fig.~\ref{fig:magMn}, where the change of the Mn magnetization $M$ (per Mn) from DFT+$U_{\rm eff}$ and eDMFT results (at $T$ = 232 K) are shown as a function of pressure. Note that, the upturn of $M$ in the small-pressure regime ($<$ 10 GPa) in eDMFT is due to the enhancement of magnetic exchange interactions originating from increased kinetic energy scale under the pressure. 

\subsubsection{NiPS$_3$}
In the PM phase of NiPS$_3$, the MIT can be indicated not only by the gap opening in the PDOS, but also by the change of electron self-energies $\Sigma(\omega)$. In Fig.~\ref{fig:mag2}(a) and (c), the MIT can be noticed by the presence and absence of the dip at the Fermi level in their PDOS, but it is slightly unclear whether the phase at P = 30.4 GPa is an insulator due to the small but finite $e_{\rm g}$ DOS at the Fermi level due to the broadening. However, $\Im\Sigma(\omega)$, plotted in Fig.~\ref{fig:mag2}(b) and (d), show a clear difference between the two phases, because the presence (absence) of a pole at the Fermi level in the $e_{\rm g}$-$\Im\Sigma(\omega)$ signifies the presence (absence) of the Mott physics. 

In the AF-ordered phases, the gap opening is induced by the exchange splitting between the spin up and down components, {\it i.e.} $\Re\left(\Sigma_{\rm up} - \Sigma_{\rm down}\right)(\omega)$. In cases where $\Im\Sigma(\omega)$ is weak compared to $\Re\Sigma(\omega)$ and the frequency dependence of $\Re\left(\Sigma_{\rm up} - \Sigma_{\rm down}\right)(\omega)$ is small, then the eDMFT results become equivalent to the DFT+$U_{\rm eff}$ results. Fig.~\ref{fig:mag2}(e-j) present such situation, where the PDOSs shown in Fig.~\ref{fig:mag2}(e) and (h) become qualitatively equivalent to DFT+$U_{\rm eff}$ PDOS (not shown), with the exchange splitting of $\sim$ 4 eV at the Fermi level opening a gap for the $e_{\rm g}$ bands. Hence, in AF phases the MIT critical pressure is mainly determined by the $e_{\rm g}$ bandwidth and the exchange splitting $\Re\left(\Sigma_{\rm up} - \Sigma_{\rm down}\right)(\omega)$. Because the above quantities change continuously as the pressure is increased, it is not easy to point out at which pressure the MIT happens from the PDOS plots due to the presence of small broadening from $\Im\Sigma(\omega)$. At $T$ = 232K, the MIT seems to happen around 24 GPa, and this pressure does not change as $T$ is lowered to 116K. 

\begin{figure}
\includegraphics[width=0.4\textwidth]{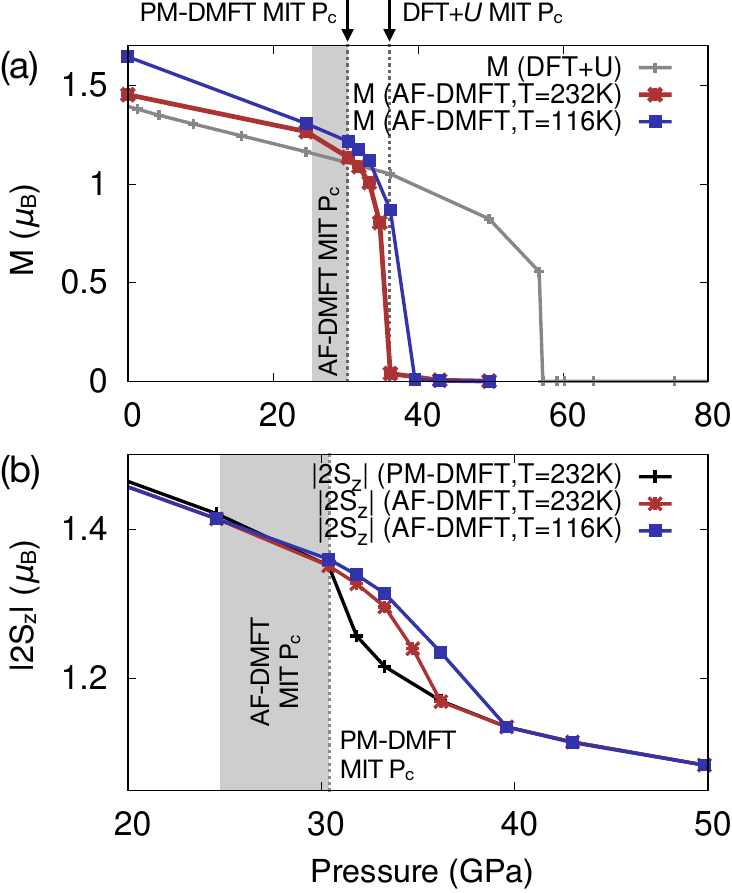}
\caption{\label{fig:mag3}(a) Evolution of Ni magnetization $M$ (per Ni) from DFT+$U_{\rm eff}$ (gray curve), AF-eDMFT at $T$ = 232 (red) and 58K (blue) as a function of pressure. The MIT critical pressures from DFT+$U_{\rm eff}$, PM- and AF-eDMFT are depicted. (b) Sizes of Ni spin moments $\vert 2S \vert$ from PM- and AF-eDMFT results. 
}
\end{figure}

Fig.~\ref{fig:mag3}(a) depicts the change of Ni magnetizations from DFT+$U_{\rm eff}$ and AF-eDMFT as a function of pressure. Note that, the MIT critical pressures are 36 and 24 $\sim$ 30 GPa for DFT+$U_{\rm eff}$ and AF-eDMFT results, as shown in the figure, but the magnetization persists within the metallic phase. The pressure where the magnetism disappears in AF-eDMFT results increases slightly from 36 to 40 GPa as the $T$ is reduced from 232K to 116K, but it does not reach 58 GPa where the magnetization disappears in DFT+$U_{\rm eff}$ results. Note that, the high-pressure anisotropic structural distortion in NiPS$_3$ happens at the pressure where the DFT+$U_{\rm eff}$ magnetization becomes zero. 

Fig.~\ref{fig:mag3}(b) shows the change of Ni local spin moment size $\vert 2S \vert$ as a function of pressure. Unlike the $\vert 2S \vert$ in the PM phase, which shows a cusp at the MIT critical pressure, $\vert 2S \vert$ in the AF phases does not show such behavior at the MIT pressure. As the pressure is increased, the AF $\vert 2S \vert$ is suppressed until the magnetization disappears and a PM metallic phase happens. Note that, the pressures that AF $\vert 2S \vert$ joins the PM $\vert 2S \vert$ curve are the points where the magnetization disappears.

\end{document}